\documentclass[preprintnumbers,  amsmath, amssymb, onecolumn,
superscriptaddress,
nofootinbib]{revtex4}
\pdfoutput=1

\usepackage{graphicx,epsfig}
\usepackage{epsfig}
\usepackage{bm}
\usepackage{amsfonts}
\usepackage{color}
\usepackage[lofdepth,lotdepth]{subfig}

\usepackage{units}

\newcommand{\bea}{\begin{eqnarray}}

\newcommand{\eea}{\end{eqnarray}}
\newcommand{\bma}{\begin{pmatrix}}
\newcommand{\ema}{\end{pmatrix}}
\newcommand{\be}{\begin{equation}}
\newcommand{\ee}{\end{equation}}
\newcommand{\beno}{\begin{equation*}}
\newcommand{\eeno}{\end{equation*}}

\newcommand{\f}{\frac}

\def\p{\partial}

\begin{document}

\title{Quantum Effects in Galileon Black Holes}

\author{George Koutsoumbas}
\email{kutsubas@central.ntua.gr}
\affiliation{Department of Physics, National Technical University of Athens, Zografou Campus GR 157 73, Athens, Greece}

\author{Ioannis Mitsoulas}
\email{ ioannis.mitsoulas@gmail.com}
\affiliation{Department of Physics, National Technical University of Athens, Zografou Campus GR 157 73, Athens, Greece}

\author{Eleftherios Papantonopoulos}
\email{lpapa@central.ntua.gr}
\affiliation{Department of Physics, National Technical University of Athens, Zografou Campus GR 157 73, Athens, Greece}

\begin{abstract}

Using the Wentzel-Kramers-Brillouin (WKB) approximation we study the formation and propagation of quantum bound states  in the vicinity of a Galileon black hole. We  show that for various ranges of the derivative coupling to the Einstein tensor, which appears  in the metric of the Galileon black hole, a Regge-Wheeler potential containing a local well  is formed. Varying the strength of the derivative coupling  we  investigate the behaviour of the bound states trapped in the potential well or penetrating the horizon of the Galileon black hole.

\end{abstract}

\maketitle

\section{Introduction}

Within the framework of Einstein's General Relativity (GR) black holes are predicted to exist and they are  extreme objects whose study is very rich and relies on
different branches of classical and quantum physics. Astrophysical black holes are supposed to be formed during the final stages of massive stars; primordial black holes may be formed in the early Universe from density inhomogeneities. Furthermore, black holes have temperature and entropy and emit radiation from the black hole horizon. The Hawking evaporation mechanism, which is a manifestation of a quantum effect in a curved spacetime (for reviews see \cite{Unruh:1976db,Crispino:2007eb}) has been studied extensively over the years. Its study is  extremely fruitful in itself  because of
the quantum gravitational effects expected to occur during the last
stages of the evaporation, when the semi-classical approach breaks down.

The detection of gravitational waves (GWs) by the Laser Interferometer Gravitational-Wave Observatory (LIGO)
Scientific Collaboration and the Virgo Collaboration
opens the window to study strong field gravitational physics
and test alternative theories of gravity \cite{Abbott:2016blz,Abbott:2016nmj,Abbott:2017vtc,Abbott:2017oio,TheLIGOScientific:2017qsa}.
 In the light
of these developments the existence of quasi-normal modes (QNMs) (for a review see \cite{Kokkotas:1999bd}) $\omega = \omega_R + i\omega_I$, with a
non-vanishing imaginary part, have become relevant, since they do not
depend on the initial conditions of the emitting objects, so they carry unique information on the few black hole parameters.

The investigation of bound
states for massive particles emitted by black holes was carried out in \cite{Grain:2007gn}. The study of this problem is  quite close to the investigation of massive QNMs
\cite{Ohashi:2004wr}, as in both cases the goal is to find the characteristic complex
frequencies allowing a massive (scalar) field to propagate in a black hole
background, while satisfying given boundary conditions at the black hole horizon
and at spatial infinity.  Nevertheless because
these boundary conditions are not the same, the physical meanings of bound states
and quasi-normal modes are very different and correspond to different energy
ranges. A systematic study was carried out in \cite{Grain:2007gn} using the formalism developed in \cite{Grain:2006dg}, of  quanta
trapped between the finite black hole potential barrier and the infinitely thick well, which prevents the particles from reaching infinity.

The investigation in \cite{Grain:2007gn}  of   quantum bound states  requires to solve
 quantum mechanically the Klein-Gordon equation in a
curved background of a test scalar field in the $4-$dimensional Schwarzschild
background using the WKB approximation. In this work we will carry out a similar analysis of quantum bound states of a scalar field around  Galileon
black holes. The Galileon black holes are hairy black holes which are the back-reacting spherically symmetric solutions of a scalar
field coupled to the Curvature \cite{Kolyvaris:2011fk,Rinaldi:2012vy,Kolyvaris:2013zfa,Babichev:2013cya}.  This coupling respects the shift symmetry, which is the characteristic property of Galileon theories. This symmetry does not allow the scalar field to have self-interacting terms. This property leads to
the fact that the derivative coupling constant $\lambda$ of the scalar field to the Einstein tensor $G_{\mu\nu}$ appears directly in the metric functions of the Galileon black hole solutions without being connected with the other parameters of the solution like the mass or the charge.

Classically this coupling has been studied extensively. The main effects of the kinematic coupling of a scalar field to the Einstein tensor
is that  gravity  influences strongly  the  propagation of the scalar field compared to a scalar field minimally coupled to gravity.
The presence of this coupling acts as a friction term in the inflationary field of the cosmological
evolution  \cite{Amendola:1993uh,Sushkov:2009hk,germani}. Moreover, it was found that at the end of
inflation there is a fast decrease of the kinetic
energy of the scalar field and this leads to a suppression of heavy particle production  as the strength of this  coupling is increased \cite{Koutsoumbas:2013boa}.
   This change of the kinetic energy of the scalar field to Einstein tensor allowed to holographically simulate the effects of a high concentration of impurities in a material \cite{Kuang:2016edj}. It has also been found \cite{Kolyvaris:2018zxl} that, for a  massive charged scalar wave kinematically coupled  to the Curvature, a trapping potential formed outside the horizon of a Reissner-Nordstr\"om black hole, and, as the strength of the coupling  was increased, the scattered wave off the horizon of the black hole was  superradiantly amplified, resulting to the instability of the Reissner-Nordstr\"om spacetime.

The aim of this investigation is to study the quantum mechanical effects of the Galileon black holes. We will consider a test wave in the vicinity of a Galileon black hole. We will show that a Regge-Wheeler potential  is formed, containing a local well, for some ranges of the derivative coupling $\lambda$  of the scalar field to the Einstein tensor. Then, varying the strength of the coupling of the scalar field to Curvature we will investigate the behaviour of the bound states trapped in the potential well or penetrating the horizon of the Galileon black hole and compare this behaviour with the SAdS case.

We note  that the test wave is not coupled to the scalar field which produces the hair to the Galileon black hole.
It understands the presence of the scalar field through the  Galileon black hole metric, in which the derivative coupling  $\lambda$ appears as a parameter. There are various local solutions of the Galileon theory. A perturbative black hole solution was discussed in \cite{Kolyvaris:2011fk}. An exact Galileon black hole solution was presented in \cite{Rinaldi:2012vy} in which however the scalar field should be considered as a particle living outside the horizon of the black hole because it blows up on the horizon. The other exact Galileon black hole solution was discussed in \cite{Babichev:2013cya} in which the scalar field is regular on the horizon but it is time dependent. In \cite{Koutsoumbas:2015ekk} the gravitational collapse is examined in this context.

The recent results on the gravitational waves \cite{Abbott:2016blz,Abbott:2016nmj,Abbott:2017vtc,Abbott:2017oio,TheLIGOScientific:2017qsa} raised some questions on the existence and stability of the scalar-tensor theories with the derivative coupling  $\lambda$ present in the theory. If we assume that the scalar field  plays the role of dark energy and drives the late cosmological expansion it has  been found \cite{Germani:2010gm} that
the propagation speed of the tensor perturbations  is different from the speed of light $c$. Due to small deviation  of the speed of GWs from the speed of light,
the dark energy models that predict $c_{gw}\neq c$ at late cosmological times are severely constrained \cite{Sakstein:2017xjx,Ezquiaga:2017ekz}. This coupling introduces a mass scale in the theory $M^2=1/\lambda$ and taking under consideration the early and late cosmological evolution the derivative coupling is bounded to be $10^{15}\text{GeV}\gg M \gtrsim 2\times 10^{-35}$GeV \cite{Gong:2017kim}. Local solutions of scalar-tensor theories with a derivative coupling can exist in various energy scales and their  stability  has been discussed in the literature \cite{Kolyvaris:2018zxl,Minamitsuji:2014hha,Yu:2018zqd}.  On the other hand,  it has been shown in \cite{Babichev:2017lmw} that a disformal transformation of the original metric to a physical metric  guarantees the validity  and the   stability of Galileon black holes. The derivative coupling appears as a parameter in the physical metric \cite{Babichev:2017lmw} and our analysis is also valid in the transformed solution defining a modified Regge-Wheeler potential.

An interesting question is whether we expect any clear observational differences of the hairy scalar-tensor black holes, such as the Galileon black holes,
from  GR black holes. If the scalar field is fixed then the background solutions are the same as in GR and possible differences can be seen only
if we perturb space-time. However, if the hairy scalar-tensor theories are mathematically and physically consistent, we expect clear observational differences from GR  black  holes.    Similar  deviations  from  GR  can  be  expected from astrophysical black holes and  neutron  stars. For
example  calculating the quasi-normal modes or possible  gravitational  and  scalar  radiation  from  binaries  in  the  case  of  scalar-tensor theories deviations from GR can be detected. Therefore it is important to study compact strongly gravitating objects in modified gravity theories. Quantum effects could also be important in the early stages of structure formation.

The work is organized as follows. In Section \ref{regge} we review the formation of the Regge-Wheeler potential for a test scalar field outside a black hole. In Section \ref{reggeh} we discuss the formation of quasi-bound states and their behaviours outside a Galileon black hole, while the last Section contains our conclusions.

\section{The Regge-Wheeler potential}
\label{regge}

To study quantum bound states around a  black hole  one has to  solve
relativistic quantum mechanical equations in a
curved background. To show that
those states do exist, the Klein-Gordon equation in the black hole background
has to exhibit a radial potential containing a local well for given
ranges of black hole horizon radii.

The Klein-Gordon equation of motion for a massless scalar field $\Phi$  in a spherically symmetric spacetime with metric
\be ds^2 = -h(r) dt^2 +\f{dr^2}{h(r)} +r^2 d\theta^2 +r^2 \sin^2\theta d\phi^2~,\label{metr1}\ee
reads
\be \f{1}{\sqrt{-g}}\p_\mu\left[\sqrt{-g} g^{\mu\nu} \p_\nu \Phi \right]=0~. \label{kleineq}\ee
Writing $\Phi(t,r,\theta,\phi)=e^{-i D t}Y^\ell_m(\theta,\varphi)R(r)$ to split the
temporal, angular and radial parts of the field ($Y^\ell_m$ are the spherical
harmonics), the radial function $R(r)$ obeys, in a spherically symmetric space-time
background (\ref{metr1}), the equation
$$h(r) \f{d}{dr}\left( h(r) r^2 \f{dR(r)}{dr}\right)+\left[D^2 r^2-l(l+1)h(r)\right]R(r) =0~.$$
 With the substitution  $R(r)=\f{u(r)}{r}$ the radial equation above becomes
   \be r h(r) \f{d}{dr}\left(h(r) \f{d u(r)}{dr }\right) - h(r) \f{d h(r)}{dr } u(r)+\left[D^2 r^2-l(l+1)h(r)\right]\f{u(r)}{r} =0~.\label{rad1}\ee
   Introducing the tortoise coordinate $r_*$
           $$dr_*=\f{dr}{h(r)}\Rightarrow h(r)\f{d}{dr}=\f{d}{d r_*}\Rightarrow h(r)\f{d}{dr}\left(h(r) \f{d u(r)}{dr }\right) =\f{d^2 u}{d r_*^2}~,$$
    we get \be h(r) \f{d}{dr}\left( h(r) r^2 \f{dR(r)}{dr}\right) =r \f{d^2 u(r)}{d r_*^2}- h(r) \f{d h(r)}{dr } u(r)\label{h}~.\ee
so that the radial equation (\ref{rad1}) takes the form \be-\f{d^2 u}{d r_*^2}+h(r)\left[\f{l(l+1)}{r^2}+\f{1}{r}\f{d h(r)}{dr }\right] u(r)=D^2 u(r)~.\label{rad2}\ee The usual quantum mechanical
techniques can be employed in the tortoise coordinate system and using the  Chandrasekhar convention the last term of equation (\ref{rad2})  is interpreted as the squared Regge-Wheeler potential \cite{Grain:2007gn}
$$V_{RW}^2(r) \equiv h(r)\left[\f{l(l+1)}{r^2}+\f{1}{r}\f{d h(r)}{dr }\right]~,$$
so as to recover the standard
Hamilton-Jacobi equation,  and therefore the radial equation (\ref{rad2})  can be written in a Schr\"odinger-like form.

In general we shall be looking for stationary state solutions of the Klein-Gordon equation, which take the form $\Psi(\vec{r},t)=\psi(\vec{r})\exp[-i C t],$ where $C$ is in general a complex number, which will depend on $l,$ the angular momentum parameter. We will use the notation $C=\Omega-i \Gamma,\ \Gamma>0,$ where $\Omega$ will be the energy level and $\Gamma$ the corresponding bandwidth. For a given $l$ there will be several overtones, i.e. frequencies with greater $\Omega,$ which will be characterized by an integer $n.$ This fact parallels the energy spectrum of hydrogen-like atoms, where a given angular momentum is shared by more than one principal quantum numbers. Summing up, our notation will be \be C_{ln}=\Omega_{ln}-i \Gamma_{ln},\ \Gamma_{ln}>0.\ee In the numerical calculations we will use the dimensionless counterparts \be \f{\Omega_{ln}}{m_p}-i \f{\Gamma_{ln}}{m_p}\equiv \omega_{ln}-i\gamma_{ln},\ \gamma_{ln}>0.\ee

\section{Hairy Black Hole Solution of a Scalar Field Coupled to Curvature }
\label{reggeh}

In this Section we review the hairy black hole solution presented in \cite{Rinaldi:2012vy}. Consider the Lagrangian
\bea\label{lagr}
L= {m_p^{2}\over 2}R-{1\over 2}\left(g^{\mu\nu}-{\lambda\over m_p^{2}}G^{\mu\nu}\right)\partial_{\mu}\varphi\partial_{\nu}\varphi~,
\eea
where $m_{p}$ is the Planck mass, $\lambda$ a real number, $G_{\mu\nu}$ the Einstein tensor,  $\varphi$ a scalar field.  The absence of scalar potential allows for the shift symmetry $\varphi\rightarrow \varphi+$const, which is the relevant Galileon symmetry. To obtain static solutions with spherical symmetry, consider the following  metric ansatz
\bea
ds^{2}=-X(\rho)dt^{2}+Y(\rho)d\rho^{2}+\rho^{2}(d\theta^{2}+\sin^{2}\!\theta d\phi^{2})~.\label{func}
\eea
Varying the action  $S=\int d^{4}x \sqrt{g}L$  with respect to the metric and the scalar field we find \cite{Rinaldi:2012vy}
\bea
X(\rho)&=&{3\over 4}+{m_p^2 \rho^{2}\over 12 \lambda}-{2\mu \over m_{p}^{2}\rho}+{\sqrt{\lambda}\over 4m_{p}\rho}\arctan\left(m_{p}\rho\over \sqrt{\lambda}\right)~,\label{gtt}\\
Y(\rho)&=&{(m_{p}^{2}\rho^{2}+2\lambda)^{2}\over 4(m_{p}^{2}\rho^{2}+\lambda)^{2}F(\rho)}~,\label{grr}\\
\Phi^{2}(\rho)&=&-{m_{p}^{6}\rho^{2}(m_{p}^{2}\rho^{2}+2\lambda)^{2}\over 4\lambda (m_{p}^{2}\rho^{2}+\lambda)^{3}F(\rho)}~,\label{psi}
\eea
where $\Phi\equiv\varphi'$ and $\mu$ is a constant of integration that will play the role of a mass. Note that the function $F(\rho)$ in (\ref{func}) is very similar to the $g_{tt}$ component of a Schwarzschild Anti-de Sitter (SAdS) black hole, but contains extra terms depending on $\lambda$. In particular $X(\rho)$ contains a $\lambda-$dependent term, which acts like an effective cosmological constant. This solution, discussed  in \cite{Rinaldi:2012vy}, describes a genuine black hole with one regular horizon located at $\rho=\rho_{H}$ if $M>0$ and also there is a scalar field outside the horizon of the black hole given by equation (\ref{psi}). In the limit $\lambda \rightarrow \infty$ we recover the Schwarzschild solution and also in this limit from equation (\ref{psi}) we see that $\Phi$ vanishes. Therefore for a finite $\lambda$ we have deviations  from General Relativity with a scalar field outside the horizon of the black hole.

As we discussed in the introduction, the black hole solution given by the functions  (\ref{gtt}) and (\ref{grr}) is interesting because the "charge" of the scalar field, i.e. its coupling constant $\lambda$ to Einstein tensor, appears in the metric functions. Therefore, the information of how strongly the scalar field is coupled to curvature is recorded in the metric. This is not happening in other hairy black hole solutions like the MTZ hairy black hole \cite{Martinez:2004nb} or its extension \cite{Kolyvaris:2009pc}. If you consider a MTZ black hole with a self-interacting scalar field \cite{Dotti:2007cp} then the self-coupling of the scalar field is a secondary "charge"  and it is related to the other parameters of the metric functions.

Following the discussion in Section \ref{regge}, the Klein-Gordon equation (\ref{kleineq}) in the  background of the metric (\ref{func}) becomes
\be \sqrt{\f{X(\rho)}{Y(\rho)}}\f{d}{d \rho} \left(\sqrt{\f{X(\rho)}{Y(\rho)}} \rho^2 \f{d R_{ln}(\rho)}{d \rho}\right)+\left[C_{ln}^2 \rho^2 -X(\rho) l (l+1)\right] R_{ln}(\rho)=0~.\label{eqXY}\ee
We denote \be \f{d\rho}{d\rho_*}=H(\rho)\equiv \sqrt{\f{X(\rho)}{Y(\rho)}}~,\label{defrstar}\ee since this root plays the role of the function $h(r)$ of the previous Section.
 In addition we will use again the definitions of the tortoise coordinates \be d\rho_*=\f{d\rho}{H(\rho)}\Rightarrow H(\rho)\f{d}{d\rho}=\f{d}{d \rho_*}\Rightarrow H(\rho)\f{d}{d\rho}\left(H(\rho) \f{d f(\rho)}{d\rho }\right) =\f{d^2 f}{d \rho_*^2}~,\label{rhorhostar}\ee  and  equation (\ref{eqXY}), with the additional substitution: $R_{ln}(\rho)=\f{w_{ln}(\rho)}{\rho},$ becomes:
 \be -\f{d^2 w_{ln}}{d \rho_*^2}+\left[X(\rho) \f{l(l+1)}{\rho^2}+\f{X'(\rho) Y(\rho)-X(\rho) Y'(\rho)}{2 \rho Y^2(\rho) }\right] w_{ln}(\rho)=C_{ln}^2 w_{ln}(\rho)~.\label{eqR}\ee Thus we have determined the radial equation along with the potential squared \be v_l^2(\rho)=X(\rho) \f{l(l+1)}{\rho^2}+\f{X'(\rho) Y(\rho)-X(\rho) Y'(\rho)}{2 \rho Y^2(\rho) }~.\label{potential}\ee

For the numerical calculations we will use the Planck mass $m_p$ to eliminate dimensions. Thus, we use the dimensionless variables $M \equiv \f{\mu}{m_p},\ r \equiv m_p \rho,\ r^* \equiv m_p \rho^*,\ V_l^2(r) \equiv \f{1}{m_p^2} \left[v_l^2\left(\rho \rightarrow \f{r}{m_p}\right)\right],$ so that the functions $X(\rho)$ and $Y(\rho)$ of the metric become:
\be X(\rho)\rightarrow F(r)=\f{3}{4}+\f{r^2}{12 \lambda}-\f{2 M}{r}+\f{\sqrt{\lambda}}{4 r} \arctan\left(\f{r}{\sqrt{\lambda}}\right)~,\label{eqF}\ee \be Y(\rho)\rightarrow G(r)=\f{(r^2 + 2 \lambda)^2}{4 (r^2 +\lambda)^2 F(r)}~.\label{eqG}\ee On the other hand, equation (\ref{rhorhostar}) becomes: \be \f{d r}{d r_*}= \sqrt{\f{F(r)}{G(r)}}, \label{tort}\ee and the equation to be solved reads:  \be -\f{d^2 u_{ln}}{d r_*^2}+\left[F(r) \f{l(l+1)}{r^2}+\f{F'(r) G(r)-F(r) G'(r)}{2 r G^2(r) }\right] u_{ln}(r)=C_{ln}^2 u_{ln}(r)~,\label{eqrR}\ee where we have set: $u_{ln}(r)\equiv w_{ln}\left(\f{r}{m_p}\right).$ 
We can express the black hole mass parameter $M$ in terms of the horizon value $r_H:$ \be F(r_H)=0\Rightarrow M=\f{r_H^3 + 9 r_H \lambda + 3 \lambda^{3/2} \arctan\left[\f{r_H}{\sqrt{\lambda}}\right]}{24 \lambda}~.\ee

Then in the vicinity of the horizon the metric functions (\ref{eqF}), (\ref{eqG}) can be expanded as   \be F(r)= \left[\f{3}{2 r_H}+\f{r_H}{3 \lambda}+\f{\lambda}{4 r_H (r_H^2+\lambda)}+\f{\sqrt{\lambda} \arctan\left[\f{r_H}{\sqrt{\lambda}}\right]}{4 r_H^2}\right] (r-r_H)+O(r-r_H)^2\ee \be\equiv F'(r_H) (r-r_H)+O(r-r_H)^2~,\ee
\be G(r) \approx \f{(r_H^2+2 \lambda)^2}{4 (r_H^2+\lambda)^2 F'(r_H) (r-r_H)},\ \  dr_*\approx \f{r_H^2+2 \lambda}{2 (r_H^2+\lambda) F'(r_H)} \f{dr}{r-r_H}\ee
\be \Rightarrow r_*(r) \approx \f{r_H^2+2 \lambda}{2 (r_H^2+\lambda) F'(r_H)}  \ln|r-r_H|+K_1~,\ee while the relations \be F(r)\approx \f{r^2}{12 \lambda},\ \  G(r)\approx \f{3 \lambda}{r^2},\ \ dr_* \approx \f{6 \lambda}{r^2} dr\Rightarrow r_*(r)\approx -\f{6 \lambda}{r}+K_2~,\ee are valid at large distances. The constants $K_1$ and $K_2$ are arbitrary, but not independent. We choose $K_2=0.$ It appears that the tortoise coordinate is a small negative quantity at large $r$ and diverges logarithmically towards $-\infty$ near the horizon.

\subsection{Regge-Wheeler potential for the Galileon black hole}

\begin{figure}
\begin{center}
\includegraphics[scale=0.15]{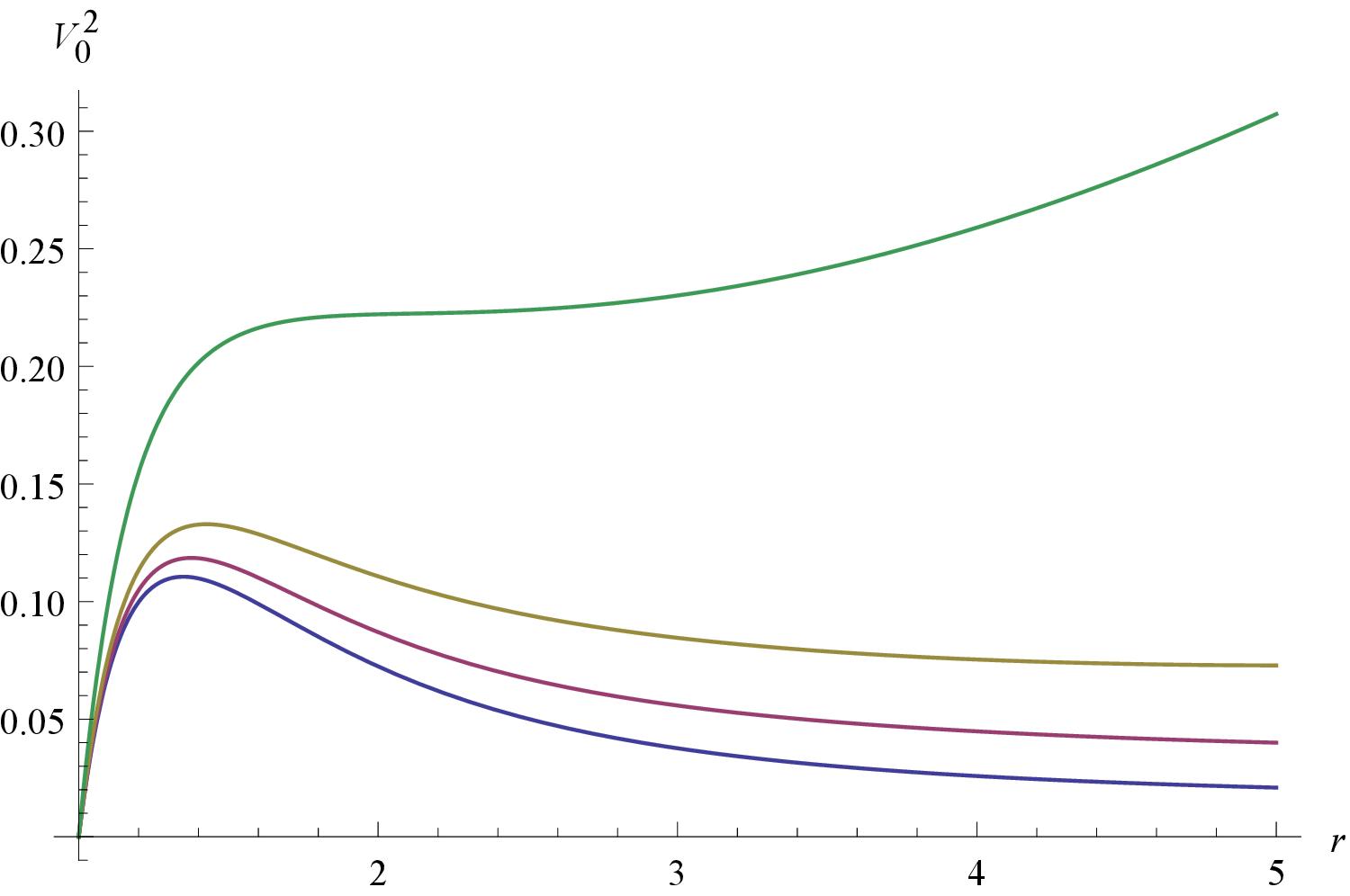}
\end{center}
\caption{Regge-Wheeler potential profiles for $l=0,\ r_H=1$ and four values of $\lambda.$ For $\lambda=50$ we get the lowest line; the remaining ones correspond to $\lambda= 20,\ 10,\ 3.$}
\label{vlambda}
\end{figure}

Let us now study the potential and point out some of its characteristics. It is actually a straightforward task, however, proceeding analytically results in too complicated expressions, so we prefer to do it numerically. In figure \ref{vlambda} we show the potential for $l=0,\ r_H=1$ and $\lambda= 3,\ 10,\ 20,\ 50.$ All the potentials diverge at infinity, due to the existence of the cosmological constant type of term. For $\lambda=50$ we get the lowest line, the potential has a peak at small $r$ and the  formed between the peak and infinity is spacious enough for (quasi-)bound states to develop. For decreasing $\lambda$  the well becomes shallower and it can support less and less bound states. When $\lambda=3$ the peak disappears and the potential cannot have bound states at all. This is a general feature of such potentials.

\begin{figure}
\begin{center}
\includegraphics[scale=0.15]{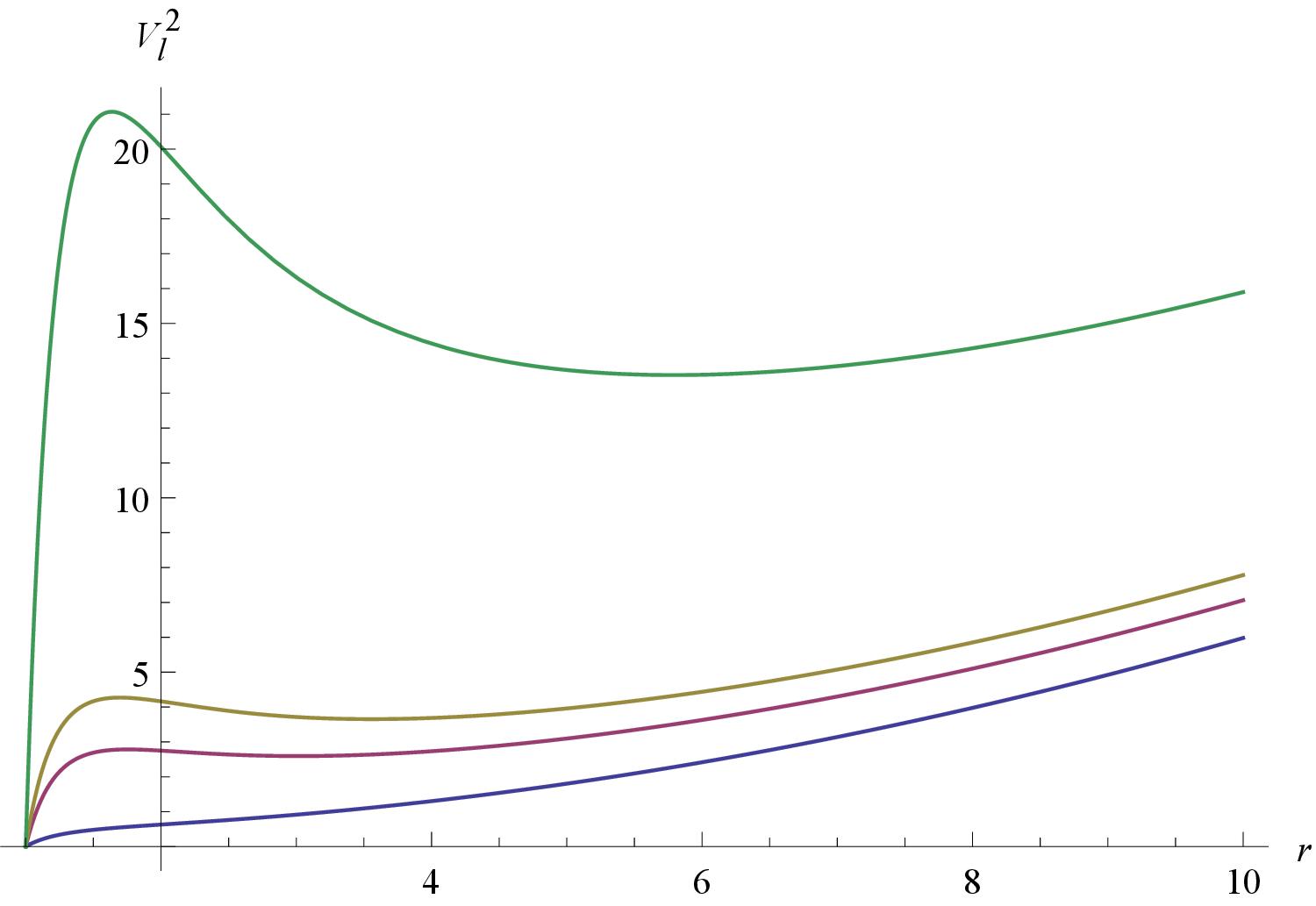}
\end{center}
\caption{Regge-Wheeler potential profiles for $\lambda=1,\ r_H=1$ and four values of $l.$ For $l=0$ we get the lowest line; the remaining ones correspond to $l= 3,\ 4,\  10.$}
\label{vl}
\end{figure}

One may wonder what is the fate of the potential for larger values of $l.$ In figure \ref{vl} we show the potential for $\lambda= 1,\ r_H=1$ and $l=0,\ 3,\ 4,\ 10.$ We have chosen a very small value for $\lambda,$ so that the potential for $l=0$ (lowest line) does not support a (quasi-)bound state. For $l=3$ we observe a small deformation, while for $l=4$  the potential develops a peak for small $r,$ which did not exist before; for $l=10$ the potential has developed a well and can support (quasi-)bound states. Thus, even for small values of $\lambda,$ one may have quasi-bound states, provided $l$ grows large enough.

\begin{figure}
\begin{center}
\includegraphics[scale=0.15]{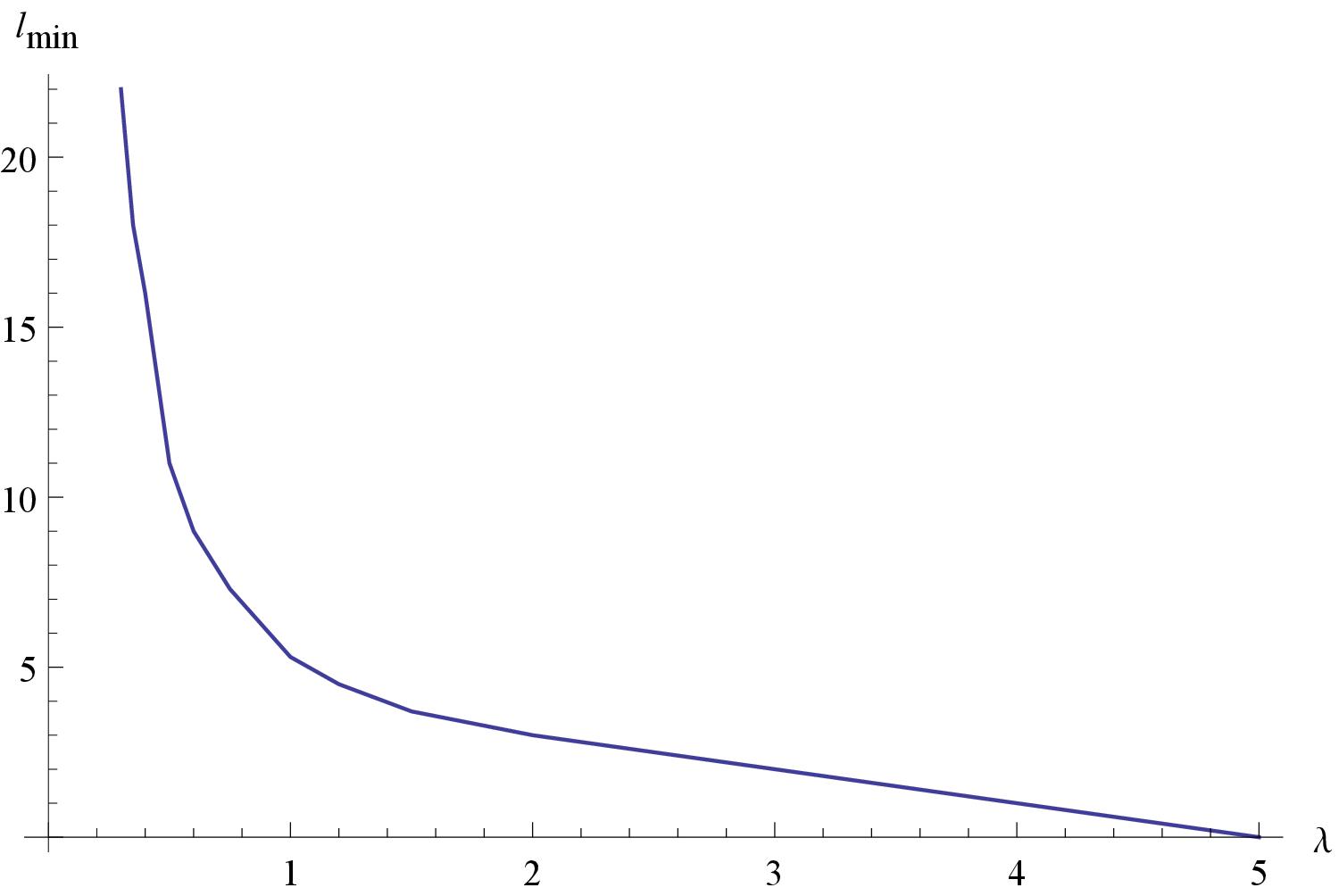}
\end{center}
\caption{Minimum value of $l,$ as a function of $\lambda$ for $r_H=1.$ For each value of $\lambda$ there exists a minimal value of $l,$ above which (quasi-)bound states are supported.}
\label{lmin}
\end{figure}

In figure \ref{lmin} we show the minimum value $l_{min}$ of $l,$ above which the potential may support bound states. The horizon parameter is $r_H=1.$ Of course, $l_{min}$ depends on $\lambda.$ For $\lambda\ge 5$ we see that $l_{min}$ is zero, that is, even for zero angular momentum the potential can support bound states. For smaller values of $\lambda$ the angular momentum value $l$ has to be bigger than some minimum value, if it is to support  bound states.

\begin{figure}
\begin{center}
\includegraphics[scale=0.15]{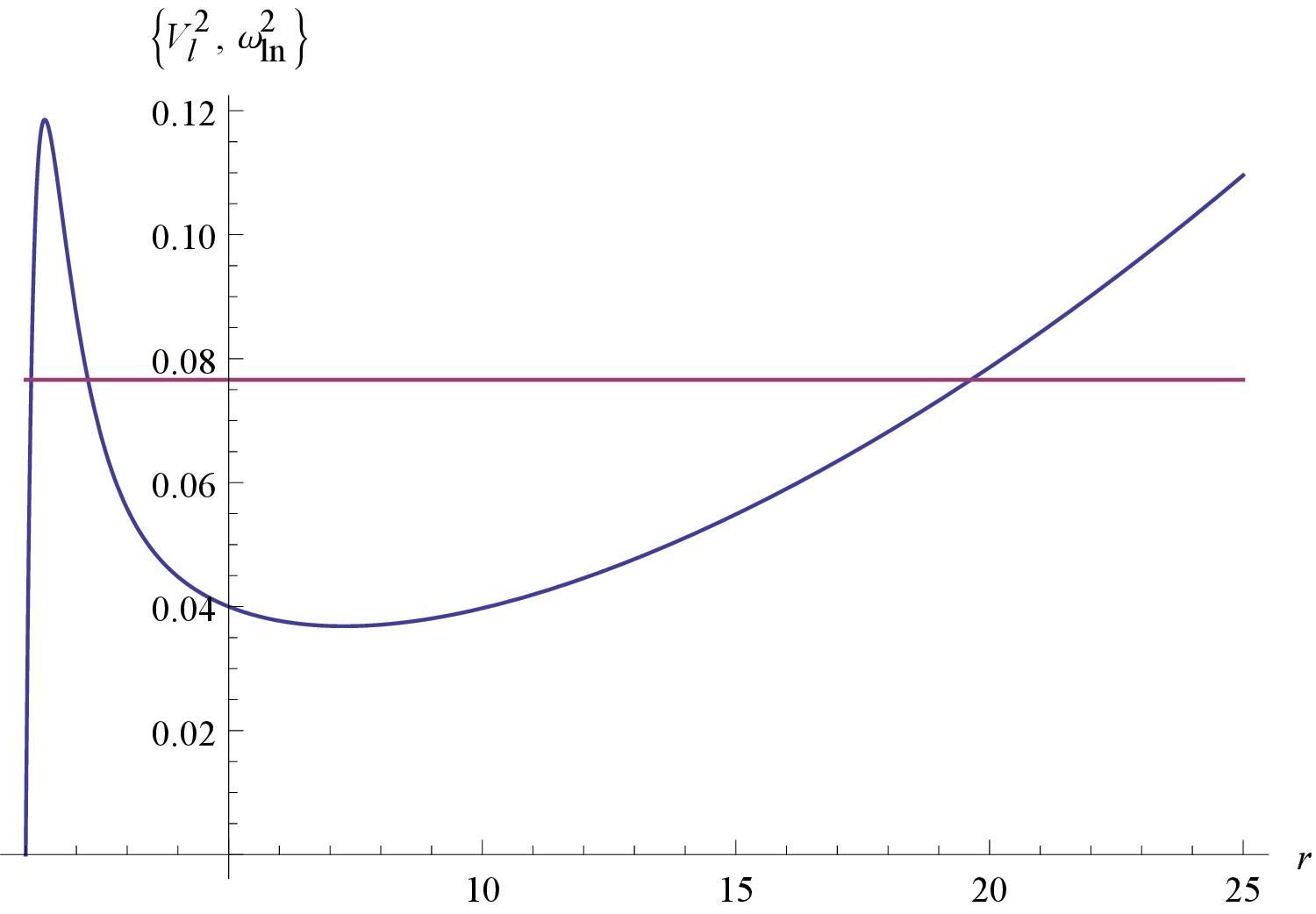}
\end{center}
\caption{A typical profile for the Regge-Wheeler potential with a generic value of $\omega_{ln}^2.$ We have chosen $r_H=1,\ \lambda=20,\ l=0,\  n=0,\ \omega_{00}^2=0.0766.$ The various points of special interest lie at $ r_-=1.108,\ r_{max}=1.376,\ r_+=2.227,\ r_0=19.634.$}
\label{pg}
\end{figure}

Let us now describe the calculating procedure for the dimensionless energies $\omega_{ln}$ and bandwidths $\gamma_{ln}.$ In figure \ref{pg} we show a typical form of the Regge-Wheeler potential $V_{l}^2(r)$ along with some value for $\omega_{ln}^2.$

The Regge-Wheeler potential has three turning points,  where the value of $\omega_{ln}^2$ cuts the potential graph of $V_l^2(r)$ and which are denoted by $r_H< r_- < r_+ < r_0.$ Since we are looking for the solution of equation (\ref{eqrR}), the independent variable is $r_*.$ Referring to the definition (\ref{tort}) we can easily see that $\f{dr}{d r_*}$ is always positive, so that the inequalities $r_H< r_- < r_+ < r_0$ yield $r_{*}(r_H)<r_{*}(r_-)<r_*(r_+)<r_{*}(r_0).$  In the sequel we will also use the notations: $r_H^*\equiv r_*(r_H),\ r^*_-\equiv r_*(r_-),\ r^*_+\equiv r_*(r_+),\ r^*_0\equiv r_*(r_0).$ Thus the horizon and the turning points divide the real $r_{*}$ axis into four regions, in which the WKB wave functions read
\be u_{ln}(r^*)=\left\{\begin{array}{ll}\f{N}{\sqrt{p(r_*)}} \left(\exp[i \int^{r^*} dx p(x)] - i \exp[-i \int^{r^*} dx p(x)]\right)& ,r^{*}_H<r^*<r^{*}_-,\\ \f{N}{\sqrt{q(r_*)}} \exp[-\f{i \pi}{4} - \int_{r_{-}^*}^{r^*} dx q(x)] & ,r^*_- <r^*<r^*_+,\\ \f{N}{2 \sqrt{p(r_*)}} \exp\left[-\int_{r_{-}^*}^{r_{+}^*} dx q(x)\right] \left(\exp[i \int^{r^*} dx p(x)] - i \exp[-i \int^{r^*} dx p(x)]\right)& ,r_+^*<r^*<r^*_0,\\  \f{N}{2 \sqrt{q(r_*)}} \exp\left[-\int_{r_{-}^*}^{r^*_+} dx q(x)\right] \exp[-\f{i \pi}{4} - \int_{r_0^*}^{r^*} dx q(x)] & ,r^*_0<r^*\end{array}\right.\ee In the preceding relations we have defined \be \begin{array}{ll}p(x) \equiv \sqrt{\omega_{ln}^2-V_{l}^2(x)} & ,\ \omega_{ln}^2>V_{l}^2(x)~,\\ q(x) \equiv \sqrt{V_{l}^2(x)-\omega_{ln}^2}& ,\ \omega_{ln}^2<V_{l}^2(x)~.\end{array}\ee In the previous equations $x$ is a dummy variable.

The energy levels can be calculated using the  equation \be \int_{r_+^*}^{r_0^*} d r_* \sqrt{\omega_{ln}^2-V_{l}^2(r_*)} = \int_{r_+}^{r_0} dr \f{\sqrt{\omega_{ln}^2-V_{l}^2(r)}}{H(r)}= \left(n+\f{1}{2}\right)\pi,\ \ n=0,\ 1,\ 2,\ \dots. \label{energy1}\ee The bandwidth can be found from the expression \cite{Grain:2006dg}
\be  \gamma_{ln}=\omega_{ln}\ \f{\exp\left[-2 \int_{r_-}^{r_+} dr \f{\sqrt{V_{l}^2(r)-\omega_{ln}^2}}{H(r)} \right]}{ 1+\exp\left[-\f{2 \pi p^2(r_{max})}{\sqrt{2\f{d^2 p^2(r_{max})}{d r_*^2}}}\right]}~.\label{bandwidth1}\ee

In equation (\ref{bandwidth1}) we have denoted by $r_{max}$ the value where the potential reaches its maximal value $\left.\f{d V_l^2}{dr}\right|_{r=r_{max}}=0.$

We may now change variables through $r = \bar{r} r_H,\ \lambda=\bar{\lambda} r_H^2$ and the formula (\ref{energy1}) reads \be \int_{\bar{r}_+}^{\bar{r}_0} d\bar{r} \f{\sqrt{\omega_{ln}^2 r_H^2 -V_{l}^2(\bar{r}) r_H^2}}{H(\bar{r})}= \left(n+\f{1}{2}\right)\pi~, \label{energy}\ee We may check that the quantities $\bar{V}_l(\bar{r}) \equiv V_{l}^2(r) r_H^2$ and $H(\bar{r})$ depend only on $l$ and the parameter $\bar{\lambda}\equiv \f{\lambda}{r_H^2}.$ The unknown quantity is the product $\bar{\omega}_{ln}\equiv \omega_{ln} r_H.$ The equation above may be written as \be \int_{\bar{r}_+}^{\bar{r}_0} d\bar{r} \f{\sqrt{\bar{\omega}_{ln}^2 -\bar{V}_{l}^2(\bar{r})}}{H(\bar{r})}= \left(n+\f{1}{2}\right)\pi~, \label{energy2}\ee Equation (\ref{bandwidth1}) is rewritten in a similar way. From this discussion it is obvious that one has to deal with just two parameters, namely $l$ and $\bar{\lambda},$ since $r_H$ does not enter $\bar{V}_{l}(\bar{r})$ or $H(\bar{r}).$

\subsection{Formation of quasi-bound states around the Galileon black hole}

Let us summarize our findings so far. We saw that Galileon black holes with bigger values of $\lambda$ can more easily  support (quasi-)bound states, while larger values of the angular momentum allow particles to live longer in a quasi-bound state. In this Section we will carry out a systematic investigation of the formation of quasi-bound state for $l=0$ and $l=1$ in the potential of the Galileon black hole.

\begin{figure}
\begin{center}
\includegraphics[scale=0.15]{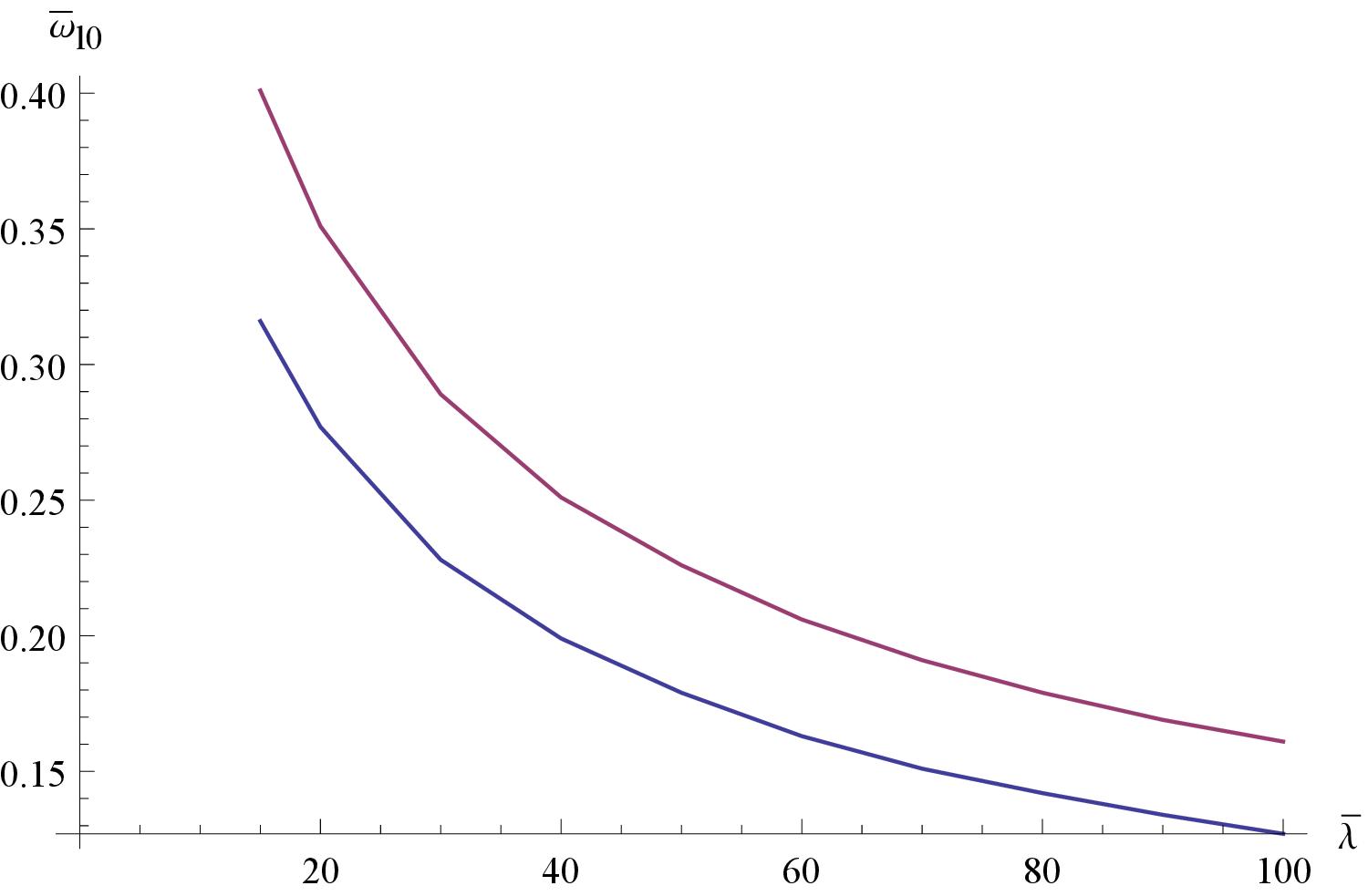}
\includegraphics[scale=0.15]{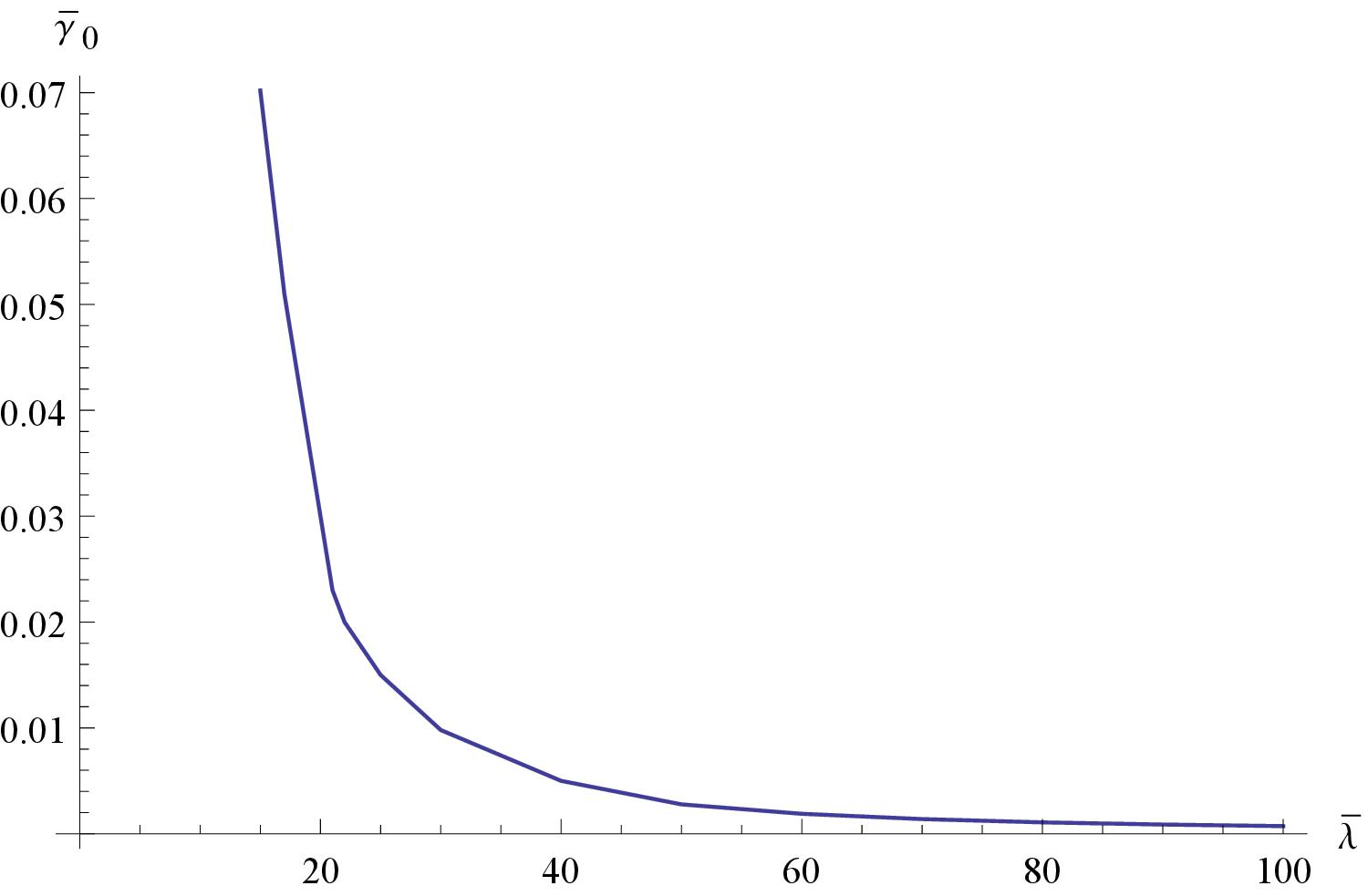}
\end{center}
\caption{In the left panel ground state energies $\bar{\omega}_{ln}\equiv \omega_{ln} r_H$ are depicted versus $\bar{\lambda}\equiv \f{\lambda}{r_H^2}$ for $n=0$ and $l=0$ (lower curves) or $l=1$ (upper curves). In the right panel we just depict
$\bar{\gamma}_{00},$ since $\bar{\gamma}_{10}$ is too small.}
\label{r1}
\end{figure}

In Fig~\ref{r1}, left panel, we display the lowest $(n=0)$ quasi-bound state energies $\bar{\omega}_{00}$ and $\bar{\omega}_{10}$ (corresponding to $l=0$ and $l=1$ respectively) versus $\bar{\lambda}.$ In the right panel we depict the corresponding quasi-bound state bandwidth $\bar{\gamma}_{00}$ (for $l=0,\ n=0$) versus $\bar{\lambda}.$ It is evident that the graphs start at a non-zero $\bar{\lambda}$ value, due to the fact that, for a given value of the angular momentum $l,$ there is always a minimal value of $\bar{\lambda},$ below which the potential can no more support (quasi-)bound states, as we have already explained. In the limit $\bar{\lambda} \to +\infty$ the Schwarzschild potential is recovered. An important remark is in place here: in the limit $\bar{\lambda}\to 0$ one does not recover the Schwarzschild potential. The black hole solution that we examine in this work lies on a branch different from the General Relativity branch, so it cannot reduce to the General Relativity solution by tuning $\bar{\lambda}.$ The curve for $\bar{\omega}_{10}$ is at higher values than the one for $\bar{\omega}_{00},$ as expected. In addition, for small $\bar{\lambda},$ the highest $\bar{\omega}_{ln}^2$ get close to the top of the potential, so that the bandwidth is large, indicating that it can easily leak out towards the horizon. This is depicted on the right panel, which shows an increase of  $\bar{\gamma}_{00}$ for small $\bar{\lambda}.$ The values of $\bar{\gamma}_{10}$ are too small, as compared against $\bar{\gamma}_{00}.$

\begin{figure}
\begin{center}
\includegraphics[scale=0.15]{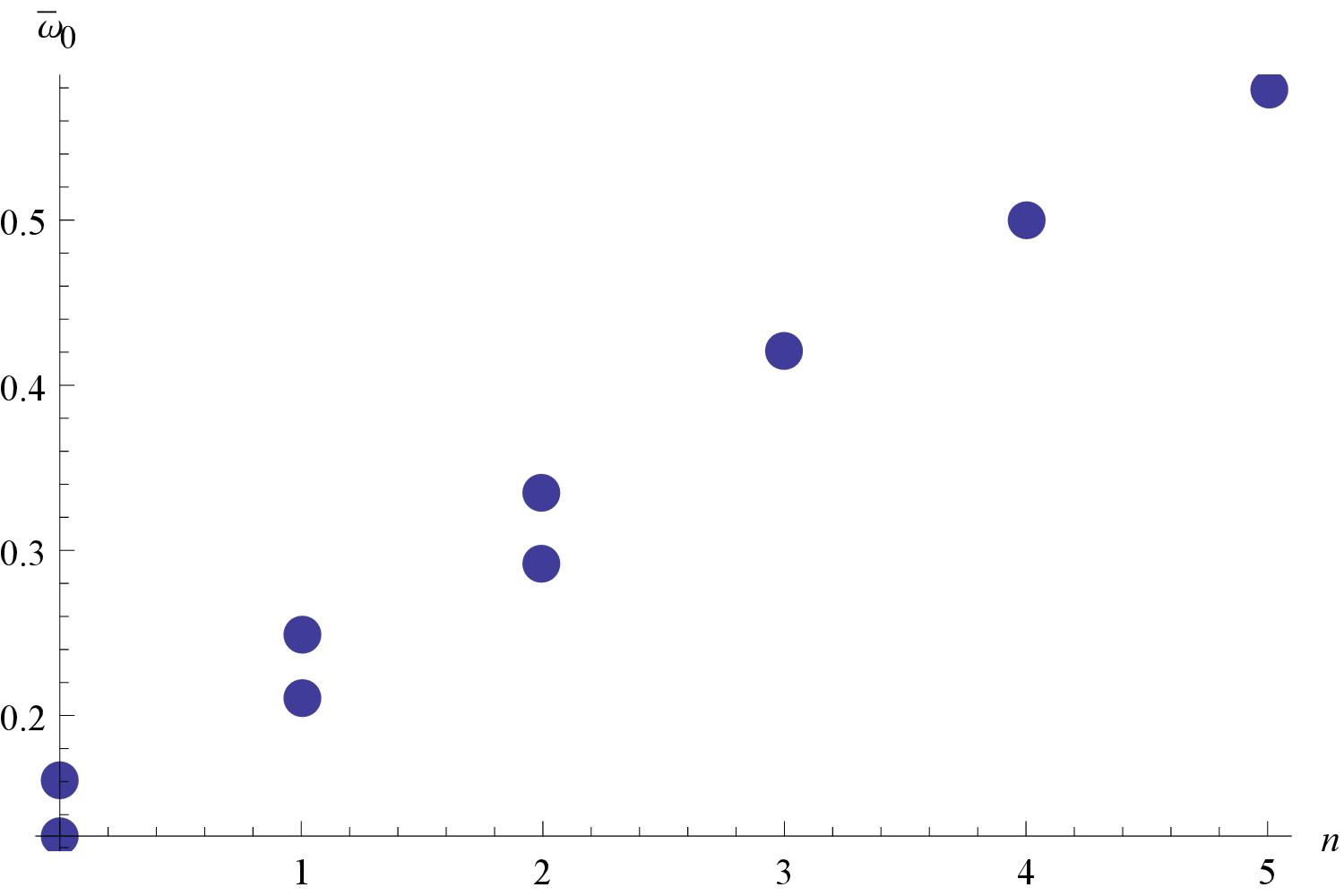}
\end{center}
\caption{Overtone energies $\bar{\omega}_{0n},\ \bar{\omega}_{1n}$ for $l=0$ (lower points) and $l=1$ (upper points), numbered by $n.$ The derivative coupling $\bar{\lambda}$ is $100$ everywhere.}
\label{r2}
\end{figure}

In Fig.~\ref{r2} we show the overtone energies associated with the angular momentum values $l=0$ and $l=1$ for fixed $\bar{\lambda}=100.$ The lower point sets represent $\bar{\omega}_{0n}$ and contain the ground state $(n=0)$ and two overtones. There exist no more overtones for $l=0$ because, at some point, the values of $\bar{\omega}_{0n}^2$ become larger than the height of the potential and there are no bound states any more. We do not show $\bar{\gamma}_{ln}$ in the graph, since we have already shown its behaviour in figure \ref{r1}. Similar remarks apply to the  $l=1$ case, however the potential is higher in this case and there exist five overtones apart from the ground state. It is evident that the variation of the values of $\bar{\omega}_{ln}$ from the ground state to the higher excitations is very regular, so one may get insight on the structure of eigenvalues by focusing on the ground states.

\begin{figure}
\begin{center}
\includegraphics[scale=0.15]{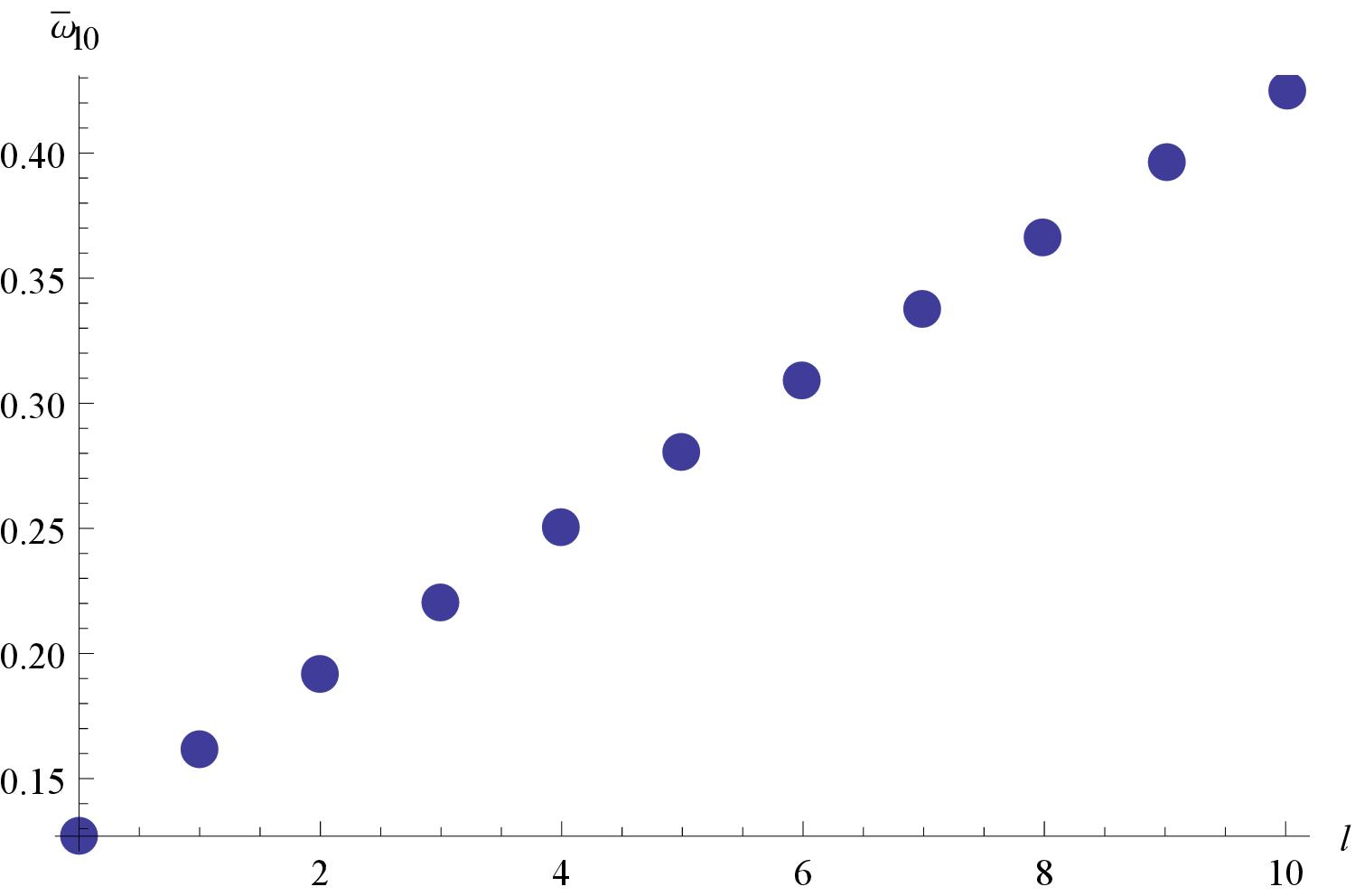}
\end{center}
\caption{Ground state energies $\bar{\omega}_{l0},\ l=0,\ 1,\ \dots,10,$ for the first eleven values of the angular momentum quantum number $l.$ The non-minimal coupling $\bar{\lambda}$ is $100$ everywhere.}
\label{r3}
\end{figure}

In Fig.~\ref{r3} we depict the ground state energies $\bar{\omega}_{l0}$ for values of the angular momentum quantum number $l$ ranging from 0 to 10. The derivative coupling $\bar{\lambda}$ is $100.$ There is nothing unexpected in this graph: the values of $\bar{\omega}_{l0}$ increase monotonically with $l.$ One may also depict the bandwidths $\bar{\gamma}_{l0}$; however, due to the great height of the potential for $l>0$ the bandwidths are too small to depict, that is the bound states are quite stable in this regime.

\section{Conclusions}

We studied the formation and the behaviour of quantum bound states outside a Galileon black hole. The Galileon black hole is a local solution of Horndeski theory and it is the result of the back-reaction  of a scalar field coupled to the Curvature for a spherically symmetric background metric. This derivative coupling to Curvature appears as a parameter in the Galileon black hole metric introducing a scale, such as the cosmological constant, and for this reason we have the formation of quantum bound states outside the horizon of the Galileon black hole.

 We  considered a test wave outside the horizon of  a Galileon black hole. We found that, for some
ranges of the coupling $\bar{\lambda},$ a Regge-Wheeler potential containing a local well is formed. Varying the strength of the derivative coupling we found that  bound states trapped in the potential well are formed.

Galileon black holes with large $\bar{\lambda}$ can support more efficiently (quasi-)bound states, while larger values of the angular momentum $l$ allow particles to live longer in a bound state. Studying the  energies and bandwidths of the bound states we found that, as $\lambda $ is increased, the overtone states become close packed and in the limit $\bar{\lambda} \to +\infty$ (the Schwarzschild limit) they form a continuum, corresponding to a continuous distribution and the absence of bound states. Moreover, the bandwidths decrease for large values of $\bar{\lambda}$ so the bound states become more and more stable while reducing $\bar{\lambda}$ renders the quasi-bound states unstable and at  some point the bound states no longer exist.

The strength of the coupling of the scalar field to Einstein tensor indicates how strongly matter is coupled to Gravity and, as we discussed  in the introduction,  it alters the kinetic properties of the scalar field propagating in a static or time-dependent space-time. An analogous behaviour is observed quantum mechanically. Compared to the Schwarzschild case we showed that bound states can be formed more easily as the strength of the derivative coupling  is increasing. We also found that, for $\bar{\lambda}\to +\infty,$ the bound states become more stable not influenced any more by a strong gravitational field. It would be interesting to extend this study to other Galileon black holes such as the black hole solution discussed in \cite{Babichev:2013cya} where the scalar field has also a time-dependence and compare it against the Schwarzschild  case.



\begin{thebibliography}{99}

\bibitem{Unruh:1976db}
  W.~G.~Unruh,
  ``Notes on black hole evaporation,''
  Phys.\ Rev.\ D {\bf 14}, 870 (1976).

\bibitem{Crispino:2007eb}
  L.~C.~B.~Crispino, A.~Higuchi and G.~E.~A.~Matsas,
  ``The Unruh effect and its applications,''
  Rev.\ Mod.\ Phys.\  {\bf 80}, 787 (2008)
  [arXiv:0710.5373 [gr-qc]].



\bibitem{Abbott:2016blz}
  B.~P.~Abbott {\it et al.} [LIGO Scientific and Virgo Collaborations],
  ``Observation of Gravitational Waves from a Binary Black Hole Merger,''
  Phys.\ Rev.\ Lett.\  {\bf 116}, no. 6, 061102 (2016)
  [arXiv:1602.03837 [gr-qc]].



\bibitem{Abbott:2016nmj}
  B.~P.~Abbott {\it et al.} [LIGO Scientific and Virgo Collaborations],
  ``GW151226: Observation of Gravitational Waves from a 22-Solar-Mass Binary Black Hole Coalescence,''
  Phys.\ Rev.\ Lett.\  {\bf 116}, no. 24, 241103 (2016)
  [arXiv:1606.04855 [gr-qc]].



\bibitem{Abbott:2017vtc}
  B.~P.~Abbott {\it et al.} [LIGO Scientific and VIRGO Collaborations],
  ``GW170104: Observation of a 50-Solar-Mass Binary Black Hole Coalescence at Redshift 0.2,''
  Phys.\ Rev.\ Lett.\  {\bf 118}, no. 22, 221101 (2017)
  [arXiv:1706.01812 [gr-qc]].



\bibitem{Abbott:2017oio}
  B.~P.~Abbott {\it et al.} [LIGO Scientific and Virgo Collaborations],
  ``GW170814: A Three-Detector Observation of Gravitational Waves from a Binary Black Hole Coalescence,''
  Phys.\ Rev.\ Lett.\  {\bf 119}, no. 14, 141101 (2017)
  [arXiv:1709.09660 [gr-qc]].



\bibitem{TheLIGOScientific:2017qsa}
  B.~P.~Abbott {\it et al.} [LIGO Scientific and Virgo Collaborations],
  ``GW170817: Observation of Gravitational Waves from a Binary Neutron Star Inspiral,''
  Phys.\ Rev.\ Lett.\  {\bf 119}, no. 16, 161101 (2017)
  [arXiv:1710.05832 [gr-qc]].

\bibitem{Kokkotas:1999bd}
  K.~D.~Kokkotas and B.~G.~Schmidt,
  ``Quasinormal modes of stars and black holes,''
  Living Rev.\ Rel.\  {\bf 2}, 2 (1999)
  [gr-qc/9909058].

\bibitem{Grain:2007gn}
  J.~Grain and A.~Barrau,
  ``Quantum bound states around black holes,''
  Eur.\ Phys.\ J.\ C {\bf 53}, 641 (2008)
  [hep-th/0701265 [HEP-TH]].

\bibitem{Ohashi:2004wr}
  A.~Ohashi and M.~a.~Sakagami,
  ``Massive quasi-normal mode,''
  Class.\ Quant.\ Grav.\  {\bf 21}, 3973 (2004)
  [gr-qc/0407009].


\bibitem{Grain:2006dg}
  J.~Grain and A.~Barrau,
  ``A WKB approach to scalar fields dynamics in curved space-time,''
  Nucl.\ Phys.\ B {\bf 742}, 253 (2006)
  [hep-th/0603042].

\bibitem{Kolyvaris:2011fk}
  T.~Kolyvaris, G.~Koutsoumbas, E.~Papantonopoulos and G.~Siopsis,
  ``Scalar Hair from a Derivative Coupling of a Scalar Field to the Einstein Tensor,''
  Class.\ Quant.\ Grav.\  {\bf 29}, 205011 (2012),
  [arXiv:1111.0263 [gr-qc]].

\bibitem{Rinaldi:2012vy}
  M.~Rinaldi,
  ``Black holes with non-minimal derivative coupling,''
  Phys.\ Rev.\ D {\bf 86}, 084048 (2012)
  [arXiv:1208.0103 [gr-qc]].



\bibitem{Kolyvaris:2013zfa}
  T.~Kolyvaris, G.~Koutsoumbas, E.~Papantonopoulos and G.~Siopsis,
  ``Phase Transition to a Hairy Black Hole in Asymptotically Flat Spacetime,''
  JHEP {\bf 1311}, 133 (2013),
  [arXiv:1308.5280 [hep-th]].

\bibitem{Babichev:2013cya}
  E.~Babichev and C.~Charmousis,
  ``Dressing a black hole with a time-dependent Galileon,''
  JHEP {\bf 1408}, 106 (2014)
  [arXiv:1312.3204 [gr-qc]].

\bibitem{Amendola:1993uh}
  L.~Amendola,
  ``Cosmology with nonminimal derivative couplings,''
  Phys.\ Lett.\  B {\bf 301}, 175 (1993)
  [arXiv:gr-qc/9302010].


\bibitem{Sushkov:2009hk}
  S.~V.~Sushkov,
  ``Exact cosmological solutions with nonminimal derivative coupling,''
  Phys.\ Rev.\  D {\bf 80}, 103505 (2009)
  [arXiv:0910.0980 [gr-qc]].

\bibitem{germani}
C.~Germani and A.~Kehagias,
  ``UV-Protected Inflation,''
  Phys.\ Rev.\ Lett.\  {\bf 106} (2011) 161302.

\bibitem{Koutsoumbas:2013boa}
  G.~Koutsoumbas, K.~Ntrekis and E.~Papantonopoulos,
  ``Gravitational Particle Production in Gravity Theories with Non-minimal Derivative Couplings,''
  JCAP {\bf 1308}, 027 (2013),
  [arXiv:1305.5741 [gr-qc]].

\bibitem{Kuang:2016edj}
  X.~M.~Kuang and E.~Papantonopoulos,
  ``Building a Holographic Superconductor with a Scalar Field Coupled Kinematically to Einstein Tensor,''
  JHEP {\bf 1608}, 161 (2016),
  [arXiv:1607.04928 [hep-th]].



\bibitem{Kolyvaris:2018zxl}
  T.~Kolyvaris, M.~Koukouvaou, A.~Machattou and E.~Papantonopoulos,
 ``Superradiant instabilities in scalar-tensor Horndeski theory,''
  Phys.\ Rev.\ D {\bf 98}, no. 2, 024045 (2018)
  [arXiv:1806.11110 [gr-qc]].

\bibitem{Koutsoumbas:2015ekk}
  G.~Koutsoumbas, K.~Ntrekis, E.~Papantonopoulos and M.~Tsoukalas,
 ``Gravitational Collapse of a Homogeneous Scalar Field Coupled Kinematically to Einstein Tensor,''
  Phys.\ Rev.\ D {\bf 95}, no. 4, 044009 (2017)
  [arXiv:1512.05934 [gr-qc]].

\bibitem{Germani:2010gm}
  C.~Germani and A.~Kehagias,
  ``New Model of Inflation with Non-minimal Derivative Coupling of Standard Model Higgs Boson to Gravity,''
  Phys.\ Rev.\ Lett.\  {\bf 105}, 011302 (2010)
  [arXiv:1003.2635 [hep-ph]].

\bibitem{Sakstein:2017xjx}
  J.~Sakstein and B.~Jain,
  ``Implications of the Neutron Star Merger GW170817 for Cosmological Scalar-Tensor Theories,''
  Phys.\ Rev.\ Lett.\  {\bf 119}, no. 25, 251303 (2017)
  [arXiv:1710.05893 [astro-ph.CO]].

\bibitem{Ezquiaga:2017ekz}
  J.~M.~Ezquiaga and M.~Zumalacárregui,
  ``Dark Energy After GW170817: Dead Ends and the Road Ahead,''
  Phys.\ Rev.\ Lett.\  {\bf 119}, no. 25, 251304 (2017)
  [arXiv:1710.05901 [astro-ph.CO]].



\bibitem{Gong:2017kim}
  Y.~Gong, E.~Papantonopoulos and Z.~Yi,
 ``Constraints on scalar-tensor theory of gravity by the recent observational results on gravitational waves,''
  Eur.\ Phys.\ J.\ C {\bf 78}, 738 (2018)
  [arXiv:1711.04102 [gr-qc]].


\bibitem{Minamitsuji:2014hha}
  M.~Minamitsuji,
  ``Black hole quasinormal modes in a scalar-tensor theory with field derivative coupling to the Einstein tensor,''
  Gen.\ Rel.\ Grav.\  {\bf 46}, 1785 (2014)
  [arXiv:1407.4901 [gr-qc]].

\bibitem{Yu:2018zqd}
  S.~Yu and C.~Gao,
  ``Quansinormal modes of static and spherically symmetric black holes with the derivative coupling,''
  arXiv:1807.05024 [gr-qc].

\bibitem{Babichev:2017lmw}
  E.~Babichev, C.~Charmousis, G.~Esposito-Farèse and A.~Lehébel,
  ``Stability of Black Holes and the Speed of Gravitational Waves within Self-Tuning Cosmological Models,''
  Phys.\ Rev.\ Lett.\  {\bf 120}, no. 24, 241101 (2018)
  [arXiv:1712.04398 [gr-qc]].

\bibitem{Martinez:2004nb}
  C.~Martinez, R.~Troncoso and J.~Zanelli,
  ``Exact black hole solution with a minimally coupled scalar field,''
  Phys.\ Rev.\ D {\bf 70}, 084035 (2004)
  [hep-th/0406111].

\bibitem{Kolyvaris:2009pc}
  T.~Kolyvaris, G.~Koutsoumbas, E.~Papantonopoulos and G.~Siopsis,
  ``A New Class of Exact Hairy Black Hole Solutions,''
  Gen.\ Rel.\ Grav.\  {\bf 43}, 163 (2011)
  [arXiv:0911.1711 [hep-th]].

\bibitem{Dotti:2007cp}
  G.~Dotti, R.~J.~Gleiser and C.~Martinez,
  ``Static black hole solutions with a self interacting conformally coupled scalar field,''
  Phys.\ Rev.\ D {\bf 77}, 104035 (2008)
  [arXiv:0710.1735 [hep-th]].














\end{thebibliography}
\end{document}